# Fundamental Mechanisms of Human Learning

**Scott E. Allen**
ORCID ID: 0000-0002-7399-3408
Department of Physics
Cornell University
Ithaca NY 14853
sea98@cornell.edu

**A. David Redish**
ORCID ID: 0000-0003-3644-9072
Department of Neuroscience
University of Minnesota
Minneapolis MN 55455
redish@umn.edu

**René F. Kizilcec** *Corresponding author*
ORCID ID: 0000-0001-6283-5546
Department of Information Science
Cornell University
Ithaca NY 14853
kizilcec@cornell.edu

# Abstract

Learning underlies nearly all human behavior and is central to education and education reform. Although recent advances in neuroscience have revealed the fundamental structure of learning processes, these insights have yet to be integrated into research and practice. Specifically, neuroscience has found that decision-making is governed by a structured process of perception, action-selection, and execution, supported by multiple neural systems with distinct memory stores and learning mechanisms. These systems extract different types of information (categorical, predictive, structural, and sequential) challenging canonical models of memory used in learning and behavioral science research by providing a mechanistic account of how humans acquire and use knowledge. Because each system learns differently, effective teaching requires alignment with system-specific processes. We propose a unified model that integrates these neuroscientific insights, bridging basic mechanisms with outcomes in education, identity, belonging, and wellbeing. By translating first principles of neural information processing into a generalizable framework, this work advances theories of skill acquisition and transfer while establishing a foundation for interdisciplinary research to refine how learning is understood and supported across domains of human behavior.

# Manuscript

## Systems of Education are Shaped by Models of Learning

Two years before the largest nuclear disaster in American history at Three Mile Island in 1979, a nearly identical plant was saved from a similar fault. The operators of the Davis Besse plant had been given scenario-based training, rather than merely rote procedural training. Thus they averted a meltdown by adaptively recognizing that a critical pressure relief valve was stuck open. Later, it was revealed that industry professionals were aware of the risks posed by the rote procedural training used elsewhere (Derivan, 2014; Malone et al., 1980; Rogovin, 1980). The sluggish response to update the training programs was a major contributing factor to the subsequent disaster (President's Commission on the Accident at Three Mile Island, 1989). Similar patterns have occurred in training and education practices in other major societal institutions including law enforcement (President's Task Force on 21st Century Policing, 2015), aviation (Federal Aviation Administration, 2001), military (Department of the Army, 2016), healthcare (Committee on Quality of Health Care in America & Institute of Medicine, 2000), and more. There are consequences for hesitating to update useful models of learning with more refined models of learning if they are available.

In applying the lessons of these events to modern education and education research, it is worth reiterating that the models of learning used to train the Three Mile Island operators were not incorrect, merely incomplete. Nuclear operators absolutely need precise knowledge of rote procedures, and they also need keen situational awareness and adaptive judgement. The training they received was essential, but it lacked refinements known at the time to be highly beneficial. Likewise, models of learning in use by today's educators and education researchers have been used to greatly benefit students' learning and wellbeing. Models such as Bloom's Taxonomy (Bloom, 1956), Deliberate Practice (Ericsson et al., 1993), Social Constructivism (Vygotsky, 1978), Cognitive Load Theory (Sweller, 1988), Self-Regulated Learning (Zimmerman, 2002), and Universal Design for Learning (Rose & Meyer, 2002) have made numerous invaluable contributions toward advancing the quality of education throughout the 20th century. These foundational models have substantially improved the practical choices that educators must make including how to present course content, how to give effective feedback, how to help students thrive amid difficulty, and how to foster a sense of safety and belonging (Ambrose et al., 2010; Bjork, 1994b; Hattie, 2008; Kapur, 2008; Shute, 2008; Walton & Cohen, 2011). They have also shaped the scientific choices of education researchers including how to measure learning outcomes, how to define core theoretical constructs, when to use qualitative versus quantitative methods, how to account for individual differences among learners, and how to develop pedagogical methods to support diverse student cohorts.

At the same time that education researchers were making these great advancements despite being unable to measure the underlying neural mechanisms for learning, neuroscientists were making great advancements towards identifying those exact mechanisms. Within the last 25 years, the field of neuroscience has produced remarkably detailed descriptions of the fundamental learning mechanisms at the level of concrete information processing operations occurring in distinctly identifiable neural systems (Churchland & Sejnowski, 1994; Dayan & Abbott, 2001; O'Keefe & Nadel, 1978; Redish, 2013; Series, 2020). These information processing operations mechanistically drive how people form habits, solve unfamiliar problems, interpret social cues, and persist through challenges . This is distinct from the best neuroscience available in the 1900's when foundational models of learning were developed, or even in the 1960's-1980's when the foundational psychological models that drove the education research of the late 20th century were developed.

Building on evidence from modern neuroscience, we start by describing learning mechanisms as information processing (Mitchell, 2023; Redish, 2013). Although the information processing in the brain depends on its physical anatomy, the neural systems governing learning and decision-making are now understood well enough to describe their function and operations without making any mention of the anatomical structures involved. Instead, we can simply describe information processing that these structures perform. Importantly, 21st century neuroscience does not diminish the vital social component of human learning or downplay the rich environmental contexts in which humans learn. Rather, it reveals how identifiable information processes in specific neural systems for perception and action-selection underlie social behaviors such as cooperation, trust repair, and belonging, making clear that the same mechanisms that govern problem solving also govern human relationships and social behavior (S. E. Allen et al., 2024; Gesiarz & Crockett, 2015; Lindquist et al., 2022; Redish et al., 2025; Thai & Lockwood, 2024). Having a concrete picture of the actual information processing that humans perform while navigating their social environments can provide a practical basis for building students' sense of safety, shared identity, and mutual trust amid the difficulties of learning. The gap that remains is that these transformative findings from neuroscience have yet to be translated into models of learning for educators and education researchers.

In this piece, we will close the gap between models of neural systems and models of learning by outlining a model which is fundamental and mechanistic. We show how these mechanisms can explain diverse behavioral outcomes — from classroom persistence and skill mastery to emotional regulation and moral decision-making — providing a framework that connects neural computation to lived human behavior. This model will benefit researchers and practitioners in many fields of human learning above and beyond formal education because it can generalize across contexts, provide a basis for understanding individual differences, accelerate new advances, and refine existing best practices. The new model we present is based on findings which have been solidly established as standard thinking in neuroscience (Gesiarz & Crockett, 2015; Rangel et al., 2008; Redish, 2013; Redish et al., 2008; van der Meer et al., 2012). Our proposed model provides novel predictions for understanding learning including novel explanations and suggestions for the iterative interaction of the learner with the environment, the role and function of multiple distinct memory systems, and the boundaries between core information processing systems involved in learning and decision-making.

# Learning is an Information Processing Phenomenon

The foundation for our model of learning is the neural process for decision-making, taken as a three-stage process of perception, action-selection, and action execution, with multiple processing components separated computationally at each stage (Redish, 2013).(Even under continuous hypotheses of this sensorimotor sequence (Cisek & Kalaska, 2010; Hayden, 2025), it is important to think in terms of the information process and information transformation.) This begins with perception processes that generate representations of the present situation. Perception is an inference about the current state of the world — the situation one finds oneself in (Friston, 2008; Gershman & Niv, 2010; Gottlieb & Balan, 2010; Pettine et al., 2023; Rao & Ballard, 1999; Redish et al., 2007). These representations of the present initiate and set the stage for the next set of processes for selecting actions. Every action that a person takes, from solving a math problem to falling in love, is governed by these action-selection systems (Redish, 2013). When an action is selected, several systems including motor and somatic control execute those actions (Cisek & Kalaska, 2010; Kawato, 1999).

Since perception forms the starting point for learning and cognition, one of the central findings of modern neuroscience essential to properly model human learning is that perception is not a passive process (Gottlieb, 2012; Gottlieb & Balan, 2010). While humans take in signals from the environment, the perception process inherently entails interactive elements. Figure 1 describes this interactive process of perception between a person and their environment. These interactions of the person with their environment include taking actions to alter the state of the environment, and most importantly, active information seeking.

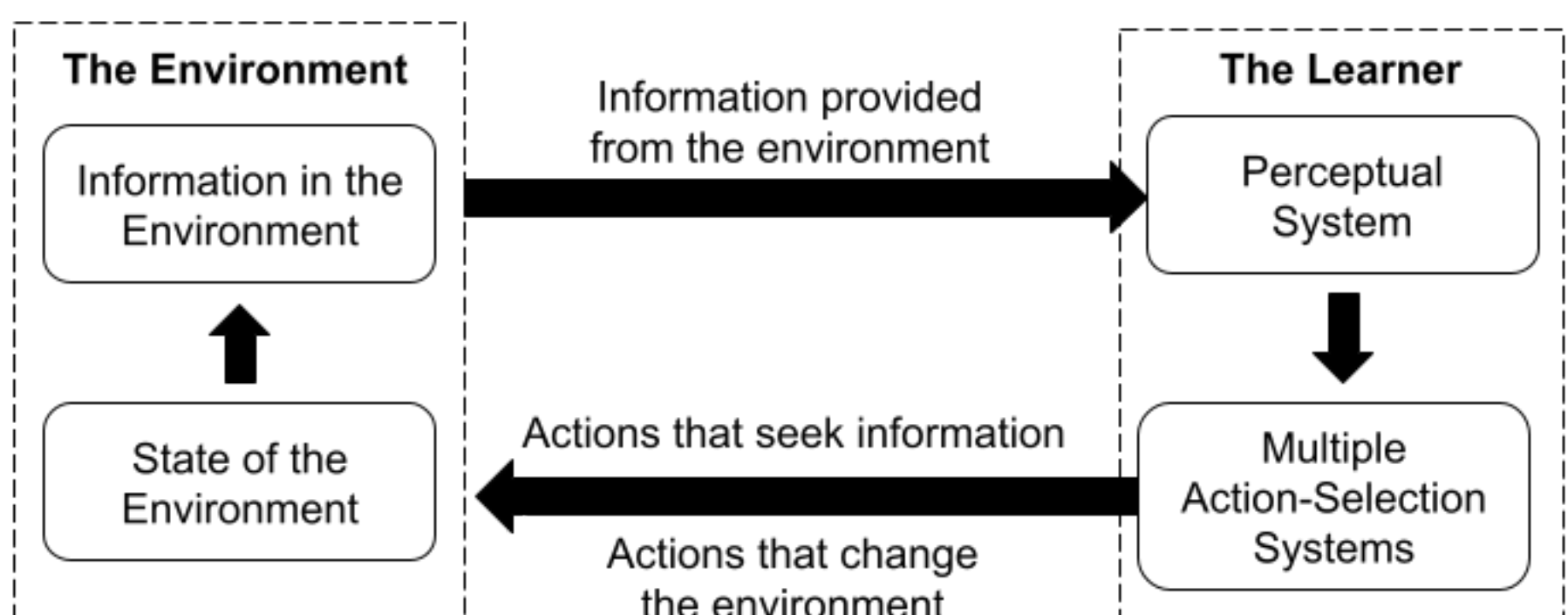

**Fig 1. Interactive Process of Perception.** Actions that the learner takes are just as much a part of the learning process as the information that the learner receives. Actions that change the environment are a part of learning, and actions that seek information are part of learning. Moreover, taking in information is an active process where incoming signals are fed through multiple layers of information processing, and perceptions of the present state of the world are actively constructed before being sent out to action-selection systems.

The next essential finding of neuroscience is an updated understanding of memory systems.[1] The nature of learning and the nature of memory are directly linked. As shown in Figure 1, interactions between the learner and the environment include inputs of information to the learner, *and* outputs of actions from the learner. Thus, the key memory systems correspond to those two broad functions: interpreting information inputs, and selecting action outputs. Just as early models of learning that described inputs as passively receiving a flow of information have been fully rejected by learning scientists and neuroscientists alike (Gottlieb, 2012; Gottlieb & Balan, 2010), recent advances in neuroscience should encourage education researchers to join neuroscientists to likewise reject the simplistic model of short-term and long-term memory in favor of new evidence-based models of memory systems (Nadel, 1994; Schacter, 2001; Squire, 1987). These new models posit distinct memory systems that are each an integral component of a corresponding information processing system, in most modern variants as one perceptual system and three action-selection systems (Gesiarz & Crockett, 2015; Rangel et al., 2008; Redish, 2013; Redish et al., 2008; van der Meer et al., 2012). In these models, memories in one system are not easily accessible by another system. A fully-specified model of learning must include how the brain conducts information-seeking actions, the multiple systems that select actions, and the memory storage that enable information processing within each system.[2] Figure 2 presents a model of human decision-making with all of these components, filling in key details for what happens in the right side of Figure 1.

---

[1] The term “memory” has varying definitions within many different disciplines. In some disciplines, it is restricted to refer only to explicit awareness of the past. Here, we use it in the broader sense to mean any information stored in the brain which is later accessed and processed. In addition to explicit memories, this includes attitudes, values, motor skills, fears, preferences, social bonds, habits, and more.

[2] Although there are other neuroscientific taxonomies, most of them entail adding additional computational processes to or variations on these three fundamental ones, such as Bayesian policy comparisons (Botvinick & Toussaint, 2012; Friston et al., 2017; Pezzulo et al., 2018), successor representations intermediate between deliberative and procedural (Momennejad et al., 2017)repeating actions independent of value (Greenstreet et al., 2025; Miller et al., 2019), or neural attractor-settling (Cisek, 2012; Hayden, 2025; X.-J. Wang, 2008). However, all of these models are based on information processing perspectives. The major implications for education and research laid out at the end of this paper remain.

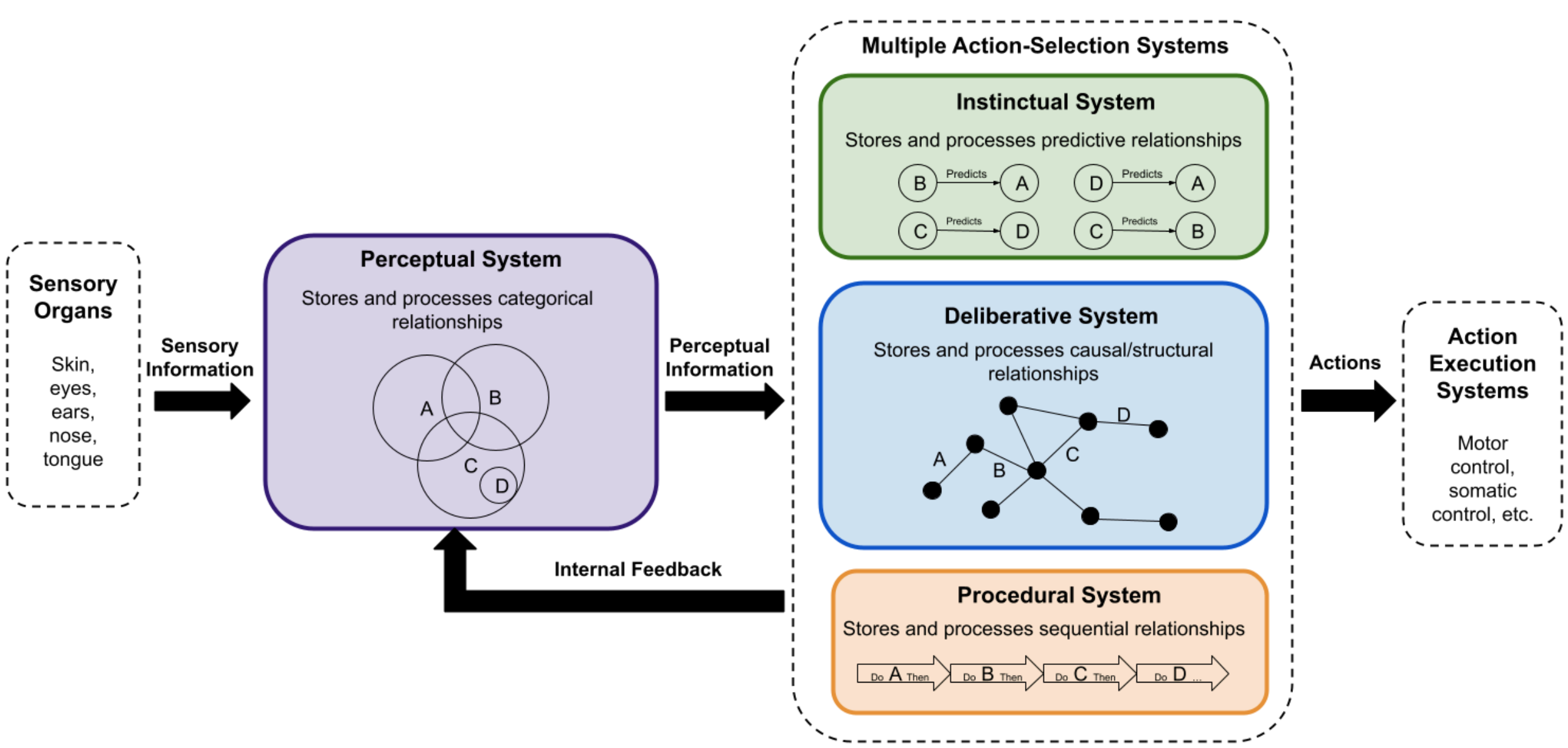

**Fig 2. Information Processing in Decision-Making Systems.** Human decision-making is a three-step process of 1) perceiving the present, 2) selecting actions, and 3) executing those actions. Cognition and learning occurs in the neural systems which perform the first two steps. The perception process categorizes the present situation and contextualizes the action-selection processes to search through a subset of possible actions relevant to that situation. Action-selection is then performed by three parallel systems. Since the action-selection systems process information in parallel, they must be modeled as separate systems. The Perceptual System and each of the three action-selection systems each have separate memory, different learning mechanisms, and interact with different information patterns of the outside environment (categorical, predictive, causal/structural, and sequential). There is an important feedback loop where action-selection systems redirect the Perceptual System to attend to features of the environment that are relevant to potential actions. This feedback is especially powerful for the Instinctual System, which produces physical responses within the body (sweaty palms, racing heart, etc) that are fed back into the Perceptual System as sensations that, along with sensations from the outside environment, are processed into a perceptual categorization the present situation.

The totality of information in human environments is enormously complex. Neural decision-making systems cope with this complexity by selectively extracting only certain types of information from the environment, and using separate systems to remember and process each type as highlighted in Figure 2. Each type of information is defined by basic relationships between one piece of information and another. In the Perceptual System, categorical relationships define the situations in human environments into overlapping groups (Gershman & Niv, 2010; McClelland et al., 2010; McClelland & Rogers, 2003; Pettine et al., 2023; Redish et al., 2007). In the Instinctual System, predictive relationships define the likelihood that one event signals the occurrence of another (Gallistel, 1990; Pavlov, 1927; Rescorla, 1988). In the Deliberative System, causal and structural relationships define maps of physical environments, and also maps of abstract ideas, which allow for many possible paths from one point to another (O'Keefe & Nadel, 1978; Redish, 1999, 2016; Tolman, 1948; Whittington et al., 2020). In the Procedural System, sequential relationships define the order in which actions occur (Dezfouli & Balleine, 2012; Graybiel, 2008). When any of these relationships between pieces of information consistently occur in the environment, it generates an *information regularity* that can be extracted from the environment by whichever neural system processes information of that type. For example, a classroom where student questions are belittled or laughed at will engender other students to become unwilling to ask questions because their Instinctual system will recognize the classroom as socially unsafe. A math student can solve a new algebra problem in a new way because their Deliberative System has extracted the causal and structural patterns they've encountered in previous problems. That same student may solve a routine problem precisely and with little effort because their Procedural System has extracted the sequence of actions they took on previous problems. The operators of the Davis Besse nuclear plant in 1977 were able to recognize the nature of the fault and avert a meltdown because their Perceptual Systems had extracted the categorical patterns from their scenario-based training environments. The processes of extracting patterns from the environment are interactive, not passive, as shown in Figure 1.

The third key finding of modern neuroscience essential for modeling learning is that information is processed separately in each decision-making system, including encoding and recall of memory (Nadel, 1994; Redish, 2013; Schacter, 2001; Squire, 1987). We emphasize that human memory is not a reservoir of information that can be arbitrarily accessed by any mental process. Each memory system is distinctly separate, and the learned memories in one action-selection system do not spontaneously transfer to the other systems (although they can be transferred through experience and consolidation processes that take time (Klinzing et al., 2019; Nadel et al., 2012; Redish, 1999; Squire, 1987)). Much of the actionable implications of this up-to-date model are consequences of the distinct boundaries between action-selection systems. Since each memory system represents information in different forms, each system has unique functionalities and unique limitations (Gesiarz & Crockett, 2015; Rangel et al., 2008; Redish, 2013; Redish et al., 2008; van der Meer et al., 2012). The Deliberative System can take creative actions in a new situation, but is often effortful in practice, variable in its execution, and often imprecise. The Procedural System can combine multiple learned skills in a precise manner, but is inflexible once learned. The Instinctual System can learn when to take genetically-inherited actions, providing fast learning, but only from a limited repertoire. To perform a learned behavior (or skill) with the functionality of a given system, a learner needs to access the memories of that specific system, which requires them to have learned these memories using the learning mechanisms unique to that system. Thus, a fully specified model of learning should respect the system boundaries between neural action-selection systems, as shown in Figure 2.

Though the Perceptual System is complex with many component sub-processes, those sub-processes occur mostly before action-selection and feed into the systems as a whole. Thus, the Perceptual System can be modeled as a single input system into the action-selection systems. Since the separate action-selection systems function in parallel, not in series, they cannot be modeled as a single system. A familiar example is that the sensory organs (eyes, ears, nose, etc), which clearly function in parallel, are always modeled as separate inputs into the Perceptual System. By analogy to the way that our sensory organs work in parallel and interact with different physical properties of the environment (electromagnetic, acoustic, chemical, etc.), our action-selection systems work in parallel and interact with different *information regularities* of the environment (see Figure 2). As we describe the systems below, we urge the reader to avoid thinking of these systems as hierarchical with certain systems governing deeper understanding or yielding better decisions. Rather, these systems perform complementary functions that support the agent in making good choices.

**The Perceptual System** finds patterns across the variations in past environments to make inferences about one's present environment (Friston, 2005; Rao, 2010; Rao & Ballard, 1999). Specifically, it uses pattern completion processes to make categorizations of present situations in the environment (Hertz et al., 1991; Hopfield, 1982; Kohonen, 1980). Pattern completion is a process through which past information is used to infer a larger picture — for example, by noticing subtle facial expressions and body language of a total stranger one can infer whether the social situation is hostile, friendly, or indifferent. The role of these processes is to produce categorizations of situations (Gershman & Niv, 2010; Grossberg, 1976; Pettine et al., 2023; Redish et al., 2007). By storing information about categorical relationships, the Perceptual System learns which features of the environment are relevant to the action-selection systems. This system governs problem recognition for cognitive skills in the same way that it processes sensory perception. Consider how a Spanish speaker can hear the difference between the phonemes /ɾ/ (a tapped 'r') and /r/ (a trilled 'rr'), distinctions that English speakers typically do not perceive. Similarly, English speakers distinguish between /ɹ/ ('r') and /l/, whereas many Japanese speakers may not initially perceive the difference (Kuhl et al., 1997; MacKain et al., 1981; Patience, 2018). Importantly, these categories are learned, not innate, and only arrive with extended experience (Groopman, 2008; G. Klein, 1999; Kuhl et al., 1997). Thus, after being well-trained, a physicist can quickly perceive important symmetries in a problem, a software engineer can identify types of syntax in code, an author can notice different narrative styles, or a teacher can pick up on subtle signs of frustration in their students.

Sensory processing is only a small part of how the Perceptual System constructs a perception of one's present environment. This entails active inattention, not mere lack of attention, to non-informative features in the environment (Gottlieb & Balan, 2010; Sims, 2003). The Perceptual System remembers these underlying patterns in order to categorize situations in the environment. Action-selection systems will use these representations to decide which actions to take pertinent to the situation. These situational categorizations also alter how experiences are evaluated as good or bad. People do not always experience a given event the same way in every circumstance. Something that is experienced as bad in one situation can be experienced as good in another. Consider experiencing chest pain during a workout versus during public speaking versus after overindulging at Thanksgiving dinner (Y. Wang et al., 2016), crying while laughing or from grief (Schmader & Mendes, 2015; Vingerhoets et al., 2012), or eye contact from a friend versus from a stranger on a dark street (Argyle et al., 1994). Evidence shows that these events are experienced much differently depending on the situation, and that these situational distinctions are learned (Groopman, 2008; G. Klein, 1999; Lakoff, 1990) . Situational categorizations also alter perception of the features in the environment (Gottlieb & Balan, 2010; Knill & Richards, 1996; Wagemans et al., 2012). Certain features that one can distinguish separately in one environment may be indistinguishable in another environment. What a person needs to recognize from the world around them while taking a physics test is very different from what they need when playing in the school orchestra.

Learners are training their Perceptual System when the learning environment requires a new scheme for classifying contexts and situations, and especially when the actions they must take are context-sensitive.

**The Deliberative System** selects actions by running a simulation of the future and evaluating the outcomes of potential actions (Doll et al., 2015; Redish, 2016). By storing information about causal and structural relationships, this system allows people to take new actions in new situations and to modify their behavior based on their immediate needs (Niv et al., 2006; O'Keefe & Nadel, 1978; Redish, 1999; Tolman, 1948). This system is involved in things like deciding between multiple job offers, or creatively solving an unfamiliar math problem. It can operate slowly for complex situations, or quickly for simple situations, but it is computationally and energetically expensive. Deliberation is physically tiring. It often utilizes heuristics, but it cannot run multiple simultaneous simulations, no matter how simple. Thus it cannot, by itself, manage the computational expense of performing complex skills with many component sub-skills. The flexible function of this system is achieved by the unique way it stores information in maps. Neural circuits which have long been known to remember maps of physical environments are now understood to also remember maps of cause-and-effect and other intricate structural relationships (Eichenbaum, 2017; Johnson & Crowe, 2009; O'Keefe & Nadel, 1978; Redish, 1999; Tolman, 1948; Whittington et al., 2020). Deliberation is thus a navigational system; it navigates physical spaces and abstract ideas alike. This system learns by mentally navigating and exploring the outcomes of possible choices, and navigating structural relationships of interrelated ideas.

Learners are training their Deliberative System when they have to think strategically about how to work through their learning exercises, and especially when a sequence of learning exercises each yields unique outcomes for any given strategy or action.

**The Procedural System** releases well-practiced actions in response to familiar features of the environment (Dezfouli & Balleine, 2012; Graybiel, 2008). Because it stores information about sequential relationships, it does not run simulations of the future or weigh the consequences of potential actions (Cunningham et al., 2021; Niv et al., 2006; van der Meer et al., 2010). This makes it computationally inexpensive during performance. Habits are often easy to execute once learned and can seem effortless. The important feature of the Procedural System is not just the speed at which it can select an action, but the ability to release complex actions reliably with many integrated component sub-skills (Doyon & Benali, 2005; Ericsson et al., 1993; Keele et al., 2003; Rosenbaum et al., 2001). This system enables routine actions like operating the steering wheel and foot pedals while driving a car, performing routine algebraic manipulations while solving a math problem, or performing routine components of everyday social situations. Even when performing these tasks slowly, the unique functionality of the Procedural System is essential in performing these multi-component tasks in a properly coordinated manner. It remembers actions, and the situations in which to release those actions. Importantly, however, this system learns through repeated practice and thus takes longer to learn and is less flexible than the Deliberative System. When an action yields an outcome that is better than expected in a given situation, the action becomes strengthened within the Procedural System. With repeated practice, actions can become fully automated as a habit (Chi et al., 1981; A. Cohen et al., 1990; Ericsson & Smith, 1991; Renkl, 2014). When multiple actions are automated and practiced together, they can be integrated into larger action chains.

Learners are training their Procedural System when they work through learning exercises that require them to practice the same action repeatedly in a high state of focus, and especially when a sequence of learning exercises contains superficially unique prompts with a common underlying pattern.

**The Instinctual System** selects actions by predicting when to release innate hereditary actions, which are genetically programmed not learned (Breland & Breland, 1961; Dayan et al., 2006; J. LeDoux & Daw, 2018; Mobbs & Kim, 2015). These actions can be complex social behaviors or simple somatic states (Gesiarz & Crockett, 2015; J. LeDoux & Daw, 2018; Rangel et al., 2008; Redish, 2013; Redish et al., 2008), but the Instinctual System only stores information about simple predictive relationships to select them (Gallistel, 1990; Rescorla, 1988). It learns which cues in which contexts predict survival-relevant situations including availability of food, presence of danger, safety, and (in humans, particularly) the degree of social belonging. Educators interested in students' sense of safety, belonging, and identity will have a particular interest in the nuances of this system. This system represents information in simple pairs of predictive relationships, and it learns via repeated recognition of those predictors (Gallistel, 1990; Pavlov, 1927; Rescorla, 1988). When a student feels unsafe and experiences unease, this is because they are perceiving combinations of cues that have predicted danger and lack of belonging in the past. In order to learn new cues, those predictions must be confounded with salient experiences of safety and respect in that moment. With exposure to consistent patterns, the Instinctual System can learn that things which predict danger and rejection in some environments do not predict danger and rejection in other environments.

Learners are using their Instinctual System when they feel comfortable enough to ask questions and make mistakes (or not), and they are training this system when they are working outside of their comfort zone. Specifically, depending on the outcomes experienced when outside of their comfort zone, this system learns to increase or decrease expectations of safety and belonging in those new contexts.

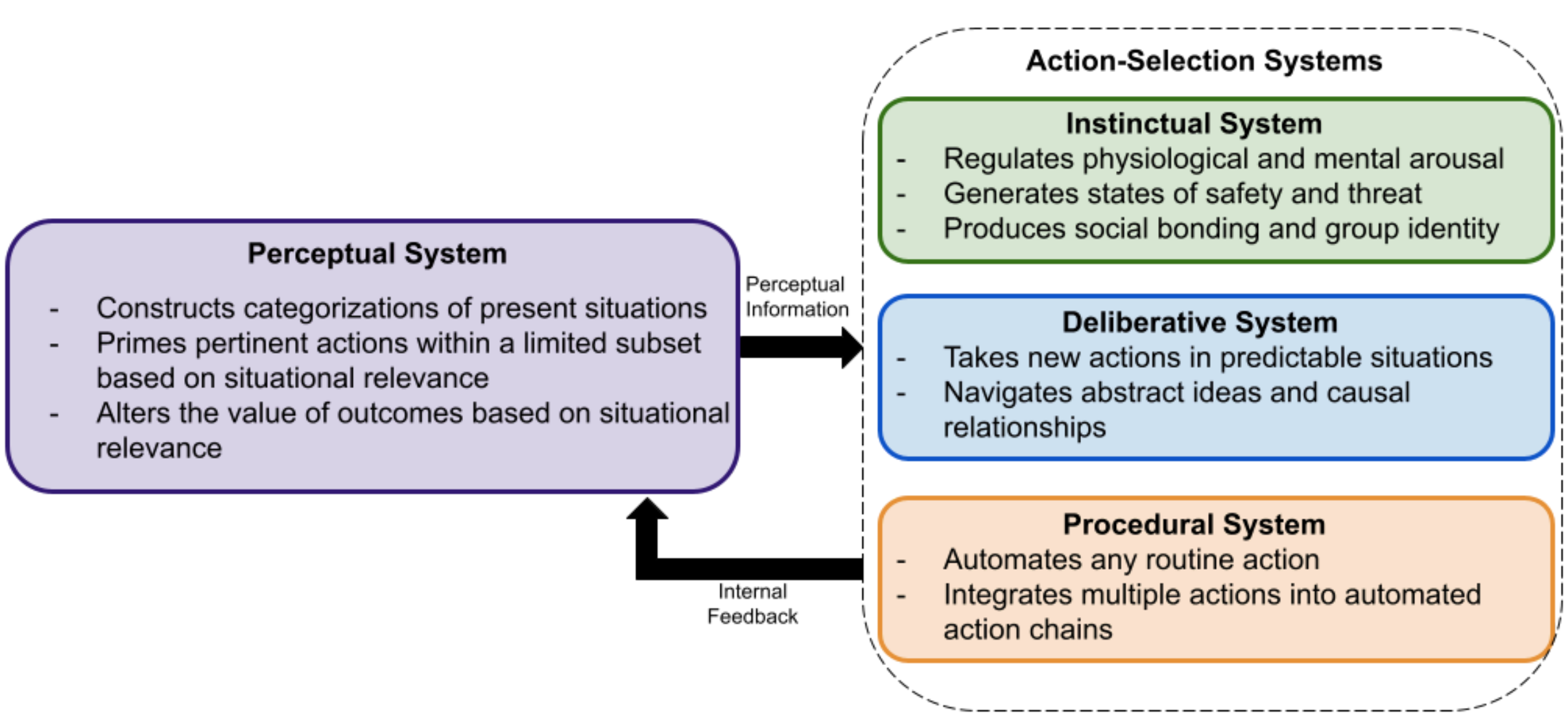

**Fig 3. Complementary Functions of Decision-Making Systems.** The system boundaries between decision-making systems are a fundamental feature of this model because each system cannot perform the functions of another system, and each system cannot learn via the learning mechanisms of any other system. For any learned behavior to utilize the beneficial functions of a given system, it must be learned by the unique learning mechanisms of that system. To overcome the limitations of any system requires support by another system. Thus, many human behaviors require multiple systems working together simultaneously. Though these systems do have identifiable interaction pathways, there is no clear evidence that any one system can monitor or regulate the others. One important interaction mechanism is the feedback loop whereby action-selection systems somewhat redirect the attention of the Perceptual System, and physical sensations resulting from the actions output by the Instinctual System are then processed and categorized by the Perceptual System.

Though all of these systems have features in common, none of these systems can completely substitute for the others. All of these systems have some fast and some slow components. All have some conscious and some unconscious components. None are purely rational or notably irrational. None are inherently selfish or altruistic, nor do they have unique goals apart from the other systems. No system specializes in any specific tasks such as driving a car, playing music, solving a math problem, or talking with a friend. Any action-selection system may output actions for any of these tasks depending on the information processing demands of the situation. Driving along a familiar route uses the same information processing operations as playing a familiar song, solving a familiar math problem, or solving a familiar social problem. Actions are controlled by the system that processes the type of information involved. Since many actions involve multiple types of information in the environment, the action-selection systems must function to complement each other. For example, driving along an unfamiliar route uses the Deliberative System to navigate, the Procedural System to operate the wheel, and the Instinctual System to maintain the proper level of mental arousal.

Though the action-selection systems are separable in that they can output behaviors independently, they work in tandem to support the functions of the others. This complementary processing can arise in situations as diverse as a student participating in a physics lesson, or a fighter pilot flying a combat mission. In either case, the Perceptual System filters out extraneous information, infers the novel and routine features of the situation, and encodes the representations of the world-state that the action-selection systems work from. The Deliberative System imagines the future and selects strategic actions for new situations. These imagined futures can redirect the Perceptual System to additional relevant features, and can allow the Instinctual System to respond to additional potential threats. The Procedural System releases routine subskills, thereby reducing the computational expense of the Deliberative System. The Instinctual System alters the somatic and affective state which provides a direct sensory input into the Perceptual System (pounding heart, sweaty palms, etc), and maintains the proper physical and mental state for the present situation.

These interactions are going to underlie all behaviors, including those of students in learning situations. Because these systems access information about the world differently (Hunt et al., 2021; MacIver et al., 2017; Mugan et al., 2024; Mugan & MacIver, 2020), educators can use classroom structures to guide how students use each of these systems to achieve their goals. Importantly, because each of these systems learns differently from different environmental conditions, educators can use the classroom environment to guide learning to specific systems. Aligning the learning goals and environment with target systems (taking into account these information processing facets) should provide better, faster, and more effective learning outcomes.

# Implications for Practice

The second half of this piece will outline some of the implications of the present model. While we illustrate our model primarily in classroom contexts, the implications extend to many societal domains. Our model suggests that aligning public health messaging with the perceptual and instinctual systems will influence the extent to which information about vaccines or nutrition is ignored, resisted, or integrated into daily routines. In organizational settings, our model suggests that structuring training to engage deliberative and procedural systems can improve teamwork, reduce costly errors, and support innovation. Our model suggests that social integration interventions that tap into the instinctual system's predictive learning can reshape experiences of threat and belonging, thereby addressing inequities in workplaces and civic life. These examples underscore that refining learning models is not only an educational challenge but also a societal imperative. For simplicity, however, we will focus on classroom education as a specific example.

Since the action-selection systems drive all human behaviors, the learning mechanisms herein should apply to all forms of human learning. Learning is the source of many human experiences including social bonding, addiction, PTSD, falling in love, moral development, repairing trust, revising beliefs, forming an identity, and behavior change of all kinds. Though this model has power to make prescriptive refinements to best practices for teaching, and we chose to use this as a familiar example, those refinements arise from a fundamental unifying framework for understanding how distinct neural systems govern the acquisition, alteration, and performance of behavior across domains as diverse as therapy, ethics, microfinance, athletics, civic engagement, and social life.

Educators must make difficult choices about how to prioritize what students should learn and how they should learn it. There are more “best practices for teaching” than any teacher could ever use, and more “good outcomes for learning” than any learner could ever pursue (Ambrose et al., 2010; Koedinger et al., 2013). There is a social and emotional balancing act as well. Push students too far beyond their comfort zone and they may get overwhelmed and discouraged, not enough and they will get bored and disengaged (Vygotsky, 1978). Using our model to parse which learning outcomes are governed by which neural system and how those individual neural systems learn best, educators can optimize, simplify, and sharpen multiple commonly used instructional strategies including (1) designing learning exercises that either integrate multiple skills together or practice them in isolation, (2) providing targeted feedback to learners, (3) using the method of contrasting cases, and (4) sequencing the order of instructional content. Our model has similar implications for the social aspect of learning. Though thinking of learning as an information processing phenomenon may seem at first to discount the social and emotional element of learning, to the contrary, we will show how understanding the complementary functions of the action-selection systems allows educators to more effectively utilize common strategies for managing the social element, including (1) minimizing extraneous difficulties, (2) collaboratively setting classroom norms, and (3) reframing the challenges of learning. We provide a brief summary of these implications in Table 1.

**Integrating different types of learning**. Our model suggests that when selecting or designing learning exercises, it may not be possible to simultaneously optimize learning in all of the different systems. Our model suggests that in some cases, different types of learning can be optimized when integrated together, but, in other cases, certain types of learning will be optimized when class time is set aside for specific types of learning and when each learning exercise is focused on a particular system.

*The Perceptual System* provides the input to contextualize the information processing required by each action-selection system to find the chosen actions relevant to the present situation. It does this by finding patterns across past experiences. There is an important feedback loop of the action-selection systems redirecting attention within the Perceptual System (see Fig 2). Though, theoretically, a student could learn the important patterns by working through a sequence of exercises where the only task is to classify the situational features — music students could practice identifying types of tempo, pilots in training could practice identifying different types of flight conditions, and physics students could practice identifying types of mathematical symmetry — these lessons are likely best optimized when integrated into decision-making tasks. For example, physics students could work through a series of exercises where they identify the type of symmetry and use that to select a solution strategy, or pilot trainees could identify the type of flight conditions and use that to identify which gauges to attend to most carefully. There is neuroscientific evidence that the Perceptual System learns faster under decision-making conditions (Bergan et al., 2005; Chi et al., 1981; Larkin et al., 1980; D. Schwartz et al., 2005). A long-standing suggestion in the education literature that learners develop expert-like problem recognition abilities via exercises that highlight the critical dimensions of a problem across multiple dissimilar contexts aligns with these hypotheses, as do the extensive data that such exercises are most effective when paired with choosing a corresponding strategy rather than practicing full solutions in detail (Chi & Glaser, 1985; Ericsson et al., 2018; Ericsson & Smith, 1991; Gick & Holyoak, 1983; Norman & Eva, 2010).

*The Instinctual System* attends to conditions such as the presence of food, danger, or social belonging that were essential to the survival of our genetic forebearers, and it releases actions relevant to those conditions. Those actions include both external actions (running from a lion) and internal actions (feeling angry at being rejected). Importantly for educational purposes, the Instinctual System actively controls a person's state of emotional, physiological, and mental arousal (Gross & Thompson, 2011; Pekrun, 2006; Pekrun et al., 2014); and all cognitive processes are affected by the continual actions of the Instinctual System. (One thinks differently when angry, scared, comfortable, or complacent (Lerner et al., 2015).) Experienced educators generally consider it ideal to allocate time specifically for social learning through ordinary practices such as icebreaker activities and other community-building exercises (D. L. Schwartz et al., 2016). While it is possible for the Instinctual System to learn in isolation from the other action-selection systems, it may not be feasible to train the other action-selection systems without considering the underlying instinctual learning that is taking place. The Instinctual System learns to recognize those states (food, danger, belonging, etc.) from cues in the environment that reliably serve as predictors of them (Dayan et al., 2006; Gallistel, 1990; Rescorla, 1988). The Instinctual system will learn whenever it is exposed to relevant predictive cues (Domjan, 1998; Mackintosh, 1974). If learning exercises for Procedural or Deliberative learning exposes students to safety or to threats (such as support or disdain for asking questions), the Instinctual system will learn the predictors of the safety or threats. The same is true of any other genetically inherited instinctual actions beyond safety and threat, such as boredom and engagement. Thus, instinctual learning is inherently integrated into all other types of learning, which implies that it is important to intentionally include instinctual learning outcomes in the design of any and all learning exercises.

*The Deliberative System* stores maps of cause-and-effect and other structural relationships. These cause-and-effect maps are generally learned through explicit identification and instruction, but the ability to navigate through these maps takes practice and skill. Students are unlikely to explore these rich knowledge structures if they follow the same line of reasoning every time and take the same action on every exercise (Jansson & Smith, 1991; Lemaire & Siegler, 1995; Mehlhorn et al., 2015). Developing rich deliberative maps is essential for adaptive problem-solving and creative thinking. By varying the situations that learners encounter and the actions that the learner takes, learners explore the space of possibilities and develop flexible knowledge (D. L. Schwartz & Bransford, 1998; Star & Rittle-Johnson, 2008). These explorations usually take the form of time-consuming discussions (Kapur, 2008). This time-consuming exploration of rich knowledge structures is in direct tension with the repetition involved in procedural learning.

*The Procedural System* stores a set of actions to release, and a set of cues for when to release them. It is optimized by using learning exercises where the learner practices the same action repeatedly. By repeating the same action, that action is strengthened until it is automated. As noted in the discussion of the Perceptual System, by varying the prompt, the Perceptual System can learn the underlying pattern across the prompts rather than the superficial features. For example, giving physics students a set of contrasting cases where they perform the same momentum analysis, but the superficial features vary between rockets, cars crashing, and hitting baseballs allows the student to learn to automate the procedure of performing momentum analysis regardless of the irrelevant features. This enables the learner to perform the practiced action in the proper circumstances. Though nobody enjoys the grind of repetitive practice (Ericsson et al., 1993), developing automated procedural action chains is essential for managing the computational load of performing complex skills (Paas & Van Merriënboer, 1994; Sweller et al., 2011). This repetitiveness does not integrate well with explicit learning exercises required for the Deliberative System (D. L. Schwartz et al., 2016). Thus, it is likely ideal to have dedicated learning exercises that are optimized specifically for procedural learning when the objective is to instill precise responses, particularly those that require sequences of actions.

**Each type of learning benefits from different kinds of feedback and contrasting cases**. Providing students with timely specific feedback is widely considered vital for students' learning (Lepper & Woolverton, 2002; C. E. Wieman, 2019). This theory predicts that each type of learning requires providing students with feedback based on the type of information that the corresponding system uses. We can informally describe these information types in terms of the "what", "why", "when", "how", and "who" of the content that students are learning. When utilizing the method of contrasting cases (D. L. Schwartz et al., 2016; D. L. Schwartz & Martin, 2004), where learners are given multiple similar exercises that draw their attention to key differences, a sequence of learning exercises can be designed to prepare students for the type of feedback corresponding to the type of learning. Thus, for example, this theory predicts that it will be better to have dedicated learning exercises that are optimized specifically for deliberative learning when the objective is to instill flexible and creative skill sets and a separate set of dedicated learning practice exercises optimized for procedural learning when the objective is to reliably perform a complex sequence of actions.

*The Perceptual System* uses information about categorical relationships to set the stage for the action-selection systems. During perceptual learning, teachers can give feedback about "what" — feedback that directs students to attend to the key features of the present situation. If students are engaged in dedicated perceptual learning, it may not be necessary to take the time to direct their attention to in-depth information about how and why to perform the skill at hand, and it may be more valuable to instead use that time to provide multiple contrasting cases and targeted feedback illustrating *what* situational features one should attend to in order to recognize those perceptual categories. The fact that perception occurs generalized across decision systems implies that one can train perception separately from the task at hand. The fact that perception is about generalization and categorization and depends on both prototypes and boundaries (Pettine et al., 2023) implies that lessons should show the range of generalization — it is likely better to show a diverse range of possibilities than it is to drill down to the prototype (Chang & Merzenich, 2003; Fisher et al., 2009; Lakoff, 1990; McClelland et al., 2002). The fact that attention is active and actions are taken to bring aspects into focus (Gottlieb & Balan, 2010) implies that such lesson plans should help students learn to use cues to bring other cues into focus. Finally, when moments of perceptual learning are integrated into other types of learning, again, feedback should direct students' attention to the important dimensions, underlying deep structure, and categorical relationships that the perceptual system can utilize, recognizing that perceptual systems are categorizing situations, not recognizing individual cues.

*The Instinctual System* uses information about predictive relationships to release hereditary actions. These hereditary actions include physical and emotional states which influence subsequent decision-making. This theory predicts that these social and emotional states released by the Instinctual System will be essential for students to recognize safety and belonging (Korpershoek et al., 2020; Pekrun, 2006), and will be key to allow for risk-taking and learning. During instinctual learning, teachers can give “who” feedback that directs students to attend to the human element, especially the students’ own internal somatic states, and the classroom as a safe space to take learning risks in. One example is reframing frustration and confusion as an ordinary part of the learning process rather than a predictor of failure (D’Mello et al., 2014; Kapur, 2008). Likewise, human-specific feelings such as shyness or fear of rejection can be constructively reframed as an ordinary part of social bonding processes rather than a predictor of exclusion. A straightforward approach of directly telling students that they belong and are expected to succeed can help foster a mental state conducive to learning (Walton & Cohen, 2007, 2011). Feedback that helps students attend to uniquely human physical and emotional states, and the role of those states as predictors can help the Instinctual system support rather than interfere with other learning systems.

*The Deliberative System* uses information about causal and structural relationships to take novel actions in predictable situations. This theory predicts that deliberative learning will be enhanced by feedback about “why” to take a given action in a given situation and about the causal consequences of those actions. Unlike perceptual learning, deliberative reasoning requires understanding consequences and being able to evaluate those consequences, which means that deliberative learning will likely be enhanced by students navigating through a chain of reasoning, considering the steps of that chain, and practicing how to evaluate outcomes as they take different paths through that mental space. A series of contrasting cases can allow students to explore why the same choice may sometimes lead to different outcomes. One such example is when medical professionals in training are given hypothetical scenarios and asked to choose a diagnosis and a treatment. Then the trainer repeatedly alters the scenario slightly, and the trainee must choose whether to make corresponding changes to their diagnosis and treatment (Elendu et al., 2024; Sheldon et al., 2023). These subtle changes allow the student to attend to *why* they made the original choice, and to construct deliberative knowledge that allows them to make correct causal reasoning in new situations that do not exactly match the scenarios they have practiced.

*The Procedural System* uses practiced sequential experiences to take precise actions in familiar situations. Procedural learning occurs by practicing the same sequence of actions repeatedly. This theory predicts that feedback about “when and how” to perform the practiced actions allows subsequent repetitions to be practiced more precisely and provide more useful procedural training than explanations of causal structure. The repetitiveness of procedural learning can be boring and exhausting (Ericsson et al., 1993). It is a common strategy to motivate students by informing them about the value of the skills they are practicing (Johansen et al., 2023; Lovett et al., 2023). One method that is both motivating and provides useful information about sequential relationships is integrating procedural learning into an authentic context — the theory predicts that the motivation can access instinctual goals and the authentic context trains the Procedural System to learn to recognize the right perceptual representations. The theory predicts that a series of contrasting cases which varies the superficial context but with a common underlying structure can provide training to both perceptual and procedural aspects and their interaction, providing information about when to release the appropriate action chain. Importantly, the theory predicts that explaining the chain logically (why one thing follows another, i.e. the *causal structure* of the chain) is not going to be helpful to training the procedural system because those are not the regularities that procedural decision-making learns from.

**Learning activities have an ideal sequence**. Just as skilled teachers strategize the sequence of various learning activities (Kim, 2018; Mager, 1961; D. L. Schwartz et al., 2016), the present model implies an ideal sequence of learning based on the information regularities of each neural system. A particular sequence is generally present in typical education systems. In kindergarten, much of the focus is on building children's attitudes about learning (Diamante et al., 2024). In post-secondary and early professional education, much of the focus is on understanding foundational concepts (LaVelle & Davies, 2021). In advanced career training, the focus shifts to efficient skills (Speck & Knipe, 2005). This same sequence can be observed over the timescale of a semester or a single lesson (Krepf & König, 2022). It begins with icebreaker activities and motivation, then progresses to exploring the conceptual structure of the content before providing students with an efficient solution. We argue that this canonical sequence works because it is transitioning from instinctual to deliberative to procedural learning, allowing each to build on the success of the previous instruction. While this general progression of instinctual to deliberative to procedural appears consistent with the implications of the present model, it often neglects perceptual learning. The action-selection systems require the perceptual system in order to function. Thus, the implication that perceptual learning should also occur early in the ideal sequence offers a key refinement to contemporary best practices for teaching and learning.

*The Perceptual System* supports the other systems, but likely learns best when practice exercises for categorizing situations are integrated with making those categorizations pertinent to important choices. These perceptual categorizations are the starting point for cognition and the starting point for learning. The model implies that perceptual learning occurs early in the ideal sequence, but continues throughout deliberative and procedural learning. Thus, we hypothesize that perceptual learning should be emphasized at regular intervals as students work their way through each of the decision systems.

*The Instinctual System* processing of social cues often strongly interacts with the other systems. Contemporary best practices for initiating a sequence of learning commonly include beginning with motivation, report-building, and other elements of instinctual learning (Bok Center, Harvard, 2025; A. K. Lane et al., 2021; Zhang, 2023). Although perceptual categorizations are the starting point for cognition and learning, including instinctual learning, the somatic states generated by the Instinctual System also support learning in the Perceptual System. Students that are socially isolated, excessively bored, or who perceive classroom experiences as threatening will experience barriers to learning in all of the other systems (perceptual, deliberative, and procedural learning). It is again important to consider feedback loops, such as when a heightened threat response causes heightened attention to threatening cues, reducing attention to the cues that the educator is trying to guide the student towards (G. L. Cohen & Garcia, 2008; Schmader et al., 2008) Beginning a course or class session by managing the social dynamics and utilizing strong motivational techniques is likely to support the learning mechanisms of the other systems.

*The Deliberative System* performs structured reasoning by running simulations based on processes for navigation. Mentally navigating to explore dead ends, inefficient solutions, unproductive paths, and unsatisfying solutions within the content knowledge can help students develop the knowledge structures needed to later perform creative decision-making in novel situations. This is difficult for both students and instructors once the students already know an efficient path to a satisfying solution. Thus, the model suggests that it is better to place deliberative learning earlier in the sequence before procedural learning, and there is emerging evidence to support this implication (D. L. Schwartz, 2024).

*The Procedural System* can perform multiple complex behaviors effortlessly and simultaneously, in part, because it does not run any simulations of the future or evaluate the outcomes from potential choices. People lose awareness of many components within complicated action chains once they are automated within the Procedural System. In other words, people function in complex environments by releasing procedural actions with little awareness of what they are doing or why they are doing it. This is often referred to as a “flow state” (Jackson & Csíkszentmihályi, 1999; Schüll, 2025) or being “in the zone” (Redish, 2013). This can directly interfere with the learning mechanisms of the other systems. It can also generate a feedback loop with the Perceptual System to cause active inattention to features of the environment that are irrelevant to choices outside of the automated action chain. In extreme cases, this is known as “professional deformation” where a person’s extensive professional training dominates their attention and thought process in situations where it should not (Merton, 1940; Polyakova, 2014). In more ordinary cases in the classroom, it can be an uphill battle to direct students’ attention away from a known efficient solution (Lemaire & Siegler, 1995; D. Schwartz et al., 2005; Star & Seifert, 2006). Thus, this model predicts that procedural learning should be saved to be the final capstone in the ideal sequence of learning.

By considering the learning mechanisms that arise from the information types used by each system, educators can craft scaffolded curricula such that learning in each system compliments and supports learning in the others.

**Social integration is a multi-system effect.** Social integration is one of the strongest enabling factors for students to thrive amid the difficulties of learning (Pascarella & Terenzini, 2005; Tinto, 1994), and our model suggests improvements to common strategies used by educators to foster a high degree of social integration for their students. Social integration improves persistence and resilience in students because it is even more than a strong rapport and sense of belonging. Social integration is a full participation in meaningful relationships with one's cohort and instructors. Thus it requires more than a lack of threatening experiences, and the multiple decision-systems model predicts that social integration is not just governed by the student's Instinctual System. While the Instinctual System is central to both producing a calm attentive state and to social bonding, full participation in meaningful social relationships likely arises from the complementary functioning of all four systems together. The effortless, automatic function of the Procedural System would allow students to perform routine help-seeking behaviors (asking questions, attending office hours, etc.) without expending their limited energy thinking about it. The flexible, imaginative function of the Deliberative System would allow students to explicitly create expectations of being treated with respect in new situations. The situation categorization function of the Perceptual System would guide students to perceive challenging or unpleasant experiences that may be signals of rejection and failure in other situations as indicators of acceptance and personal progress within the classroom context.

Since it is not possible to entirely eliminate the unpleasant challenges of learning (Bjork, 1994a; Kapur, 2008), skilled educators use many strategies to prevent these challenges from becoming barriers to social integration. Here we will address three such strategies and show how they can be further enhanced by considering the information irregularities and complementary system interactions within our model. First, we will discuss how designing learning activities into perceptual, instinctual, deliberative, and procedural types simplifies the strategy of minimizing extraneous difficulties (Immordino-Yang & Damasio, 2007; Sweller et al., 2011). Next, we will discuss how considering the complementary interaction of the Procedural and Deliberative Systems can optimize and sharpen the strategy of collaborating with students on classroom norms (Brookfield & Preskill, 2016; E. G. Cohen & Lotan, 2014). Lastly, we will discuss how considering the complementary interaction of the Perceptual System with the Instinctual System can enrich and elevate the strategy of reframing students' challenges (Bjork, 1994a; Kapur, 2008; Steele, 2010).

One basic strategy used by many educators to help students achieve social integration amid the difficulties of learning is to simply minimize the degree of unnecessary difficulties. The challenge for educators is to do so while preserving desirable difficulties (Bjork, 1994a). Along with carefully matching learning exercises to students' level of ability, educators will often parse out student difficulties by the type of barrier whether it be cognitive, emotional, cultural, physical, etc (Arnold, 2011). Our model suggests that this can be made simpler when learning exercises are parsed into perceptual, instinctual, deliberative, and procedural types. Educators can then consider for each type of learning the essential actions students must take, and thereby directly address and mitigate the challenges associated with those actions. Specifically, educators can anticipate the delayed payoff of perceptual learning, the vulnerability of instinctual learning, the confusion and frustration of deliberative learning, and the fatigue and boredom of procedural learning. Since learning in each of the four systems entails unique identifiable challenges, structuring learning exercises to target individual systems reduces the complexity of mitigating the discomfort for students while preserving the essential learning mechanisms of each type.

The same perspectives apply to essential learning behaviors beyond or alongside formal learning exercises including giving balanced participation in group discussions, asking for help, resolving conflicts respectfully, offering help to other students, and forming study groups. All of these actions are governed by the same neural systems as every other action that a person may learn, and thus depend on the same information regularities and learning mechanisms. For these actions and others, educators can systematically outline which components of social integration within their classroom are governed by each system, and intentionally provide students with learning experiences corresponding to each system.

The complementary interaction of the Deliberative System with the Procedural System has implications for how educators utilize the strategy of setting norms collaboratively with students. Common practice is to give students agency and choice over the norms, and then to provide clarity and consistency in enforcing those norms (A. K. Lane et al., 2021). Much like when providing instructional feedback, this clarity and consistency should utilize the information regularities of the neural system involved. The Procedural System automates routine actions within a situation while the Deliberative System navigates novel actions with predictable consequences. Thus when setting classroom norms, educators can consider which social actions students must take routinely and which novel social choices students must navigate. Novel actions may include resolving conflicts respectfully or developing new friendships, while routine actions may include raising a hand to ask for help or attending a study session. Any student is placed at a disadvantage if they must work through the physically tiring process of deliberating on whether to take routine these help-seeking actions. Since only the Procedural System can make routine actions effortless, and it requires repetition to learn, requiring students to take these actions early and often may be preferable to leaving it to the discretion of the students. Clarity and consistency regarding how and when to perform these routine actions is likely especially valuable. For example, in the common Think-Pair-Share format for interactive pedagogy, all students are required to participate and thus they rapidly acclimate to the prescribed expectations thereof (Guenther & Abbott, 2024). Giving students agency is beneficial, but creating norms that require students to deliberate on too many routine actions is mentally expensive and will be disproportionately expensive for some, such as first-generation students with less prior exposure to conventional norms (i.e. the hidden curriculum). In contrast, the model suggests that only the Deliberative System has the ability to navigate novel actions, but also suggests that the deliberation requires planning, which requires predictable consequences. This is a strong opportunity to collaborate with students on norms that provide students with multiple actions they might take, the reasons why they might take them, and a clear picture of the outcomes that follow from each possible action. Giving multiple options not only engages the Deliberative System, it redirects the Perceptual System to attend to the features of the situation that are relevant to that choice. But the model also suggests that being explicit (and consistent) about the consequences of these actions is important to the student's ability to deliberate and plan.

The complementary interaction of the Perceptual System with the Instinctual System has transformative implications for the common strategy of helping students reframe the challenges of learning. The model suggests that it is via this interaction that people are able to experience discomforts, difficulties, and setbacks as positive progress in appropriate circumstances. Students will have to work hard when fatigued, delay gratification, interact with new people, seek help, expose their errors, and take many kinds of assessments. Some of these experiences may be predictors of threat and rejection within many environments in the students' backgrounds. Educators often use common reappraisal strategies to normalize the struggle including: telling stories about past students who overcame challenges, sharing examples of one's own setbacks, using humor to lighten difficult moments, and reminding students that the challenge is an inherent part of the learning process (Kromka & Goodboy, 2019). Some educators even use Cognitive Behavioral Techniques (CBT) such as teaching students constructive self-talk or teaching them to reflect on their past experiences overcoming challenges (Weare & Nind, 2011). It is one thing to be aware that challenges are part of learning and to develop adaptive behaviors. It is another thing to legitimately experience struggles as positive progress. This model suggests that discomfort does not always yield negative associations because the Perceptual System and the Instinctual System interact to use discomfort as a predictor of positive or negative outcomes depending on the situation. It is obviously preferable to make learning a pleasure whenever possible, but an opportunity arises when the unpleasant aspects occur at foreseeable moments. When the Insinctual System responds to predictors of danger, and then those predictions are repeatedly confounded by real outcomes that are safer than expected in a given situation, this model predicts that two forms of learning should occur: the Perceptual System should generate a new classification for the situation, and the Instinctual System should learn the predictors of safety and progress within that new situation. Our model suggests that guidance to students can exploit this powerful complementary system interaction if it helps students attend to what makes the situation different from past situations, it pairs the the experience with clear messages about acceptance and progress, it helps students attend to their internal somatic state as the predictor of progress, and if this supportive guidance is given in the moment that the struggles occur. This perceptual reframing of the situation can also yield strong social bonds if the key differentiating feature of the situation is the presence of supportive instructors and supportive fellow students who then become part of the student's identity group within the Instinctual System.

The social and emotional outcomes that educators desire to foster in their students exist as the basic functions of neural decision-making systems; the same systems that learn the skills of the course content. These neural systems perform those basic functions via information processing. Drawing on up-to-date evidence on how these four systems perform complementary information processing, educators of all kinds can make substantial improvements to the social and the intellectual aspects of learning.

**Table 1.** Overview of neural systems and implications for practice. Columns represent distinct neural information-processing systems, while rows describe their functions, and actionable implications for instruction.

| **Function and Roles of Each Decision-Making System** | | | | |
|---|---|---|---|---|
| | **Instinctual** | **Perceptual** | **Deliberative** | **Procedural** |
| **General Functionality of each Neural System** | Maintaining proper physical and mental state | Activating relevant knowledge in the right circumstances | Navigating novel decisions in predictable situations | Performing complex routine actions with high precision and low effort |
| **Role in Academic Problem Solving** | Maintaining a state of mental arousal and curiosity | Performing problem recognition | Creative or imaginative problem solving | Automating routine components of complex problems |
| **Role in Social Integration** | Maintaining a somatic state of safety and belonging | Experiencing unpleasant challenges as positive | Imagining future events positively amid uncertainty | Performing help-seeking behaviors as habits |
| **Implications for Optimizing Learning in Each Decision-Making System** | | | | |
| **Type of Instructional Feedback** | Which internal states are predictors of success and belonging | What features of the problem indicate an ideal solution strategy or action | What are the underlying causal and structural relationships | How to perform each action correctly |
| **Type of Contrasting Cases** | Experiences that are good in one situation but bad in some others | Contrasting relevance to a decision | Contrasting outcomes from a given choice | Contrasting superficial features |
| **Type of Practice** | Stretching beyond one's comfort zone | Categorizing situations | Navigating knowledge structures | Precise repetition of actions |
| **Sequencing of Learning Exercises** | First | Second | Third | Last |
| **Isolation of Learning Exercises** | Isolated exercises and integrated into all other exercises | Integrated with the initial choices, but without in-depth practice of other systems | Isolated from procedural learning exercises | Isolated from deliberative learning exercises |

# Implications for Research

**Fundamental learning mechanisms can generate causal hypotheses and guide research design**. First principles of computation theory demand that the way that information is represented (categorical, sequential, structural, predictive, etc.) governs how it can be learned, stored, accessed, and processed (Cormen et al., 1990; Lieder & Griffiths, 2019; Simon, 1955). Incorrect or incomplete assumptions about the brain's information processing underlying the fundamental learning mechanisms can compromise the interpretation of research findings (Howard-Jones, 2014). Rigorous research on human learning should therefore account for learning mechanisms since they are well-understood down to the fundamental level of concrete information processing operations taking place. For illustration, consider the example of a large-scale study comparing the benefits of immediate feedback to delayed feedback (Fyfe et al., 2021). Since none of the neural systems are inherently fast or slow and there are no "fast" or "slow" types of information, the immediacy of the feedback is only likely to correlate with student learning if it is mediated by the information processing mechanisms of one of the systems. Without distinguishing effects by the type of learning attempted, the type of feedback provided, or whether students have opportunity to incorporate that feedback into subsequent practice, the results are expected to be variable and highly context dependent (indeed, the study yielded a high-variance null result). Our model suggests that the assessment needs to be aligned with the type of intervention: if the feedback promoted deliberative learning while the assessments measured procedural learning (or vice versa), our model suggests that the study would underestimate the students' learning gains. When selecting independent variables, outcome measures, mediators, and moderators, we recommend separating the information into predictive, sequential, cause-effect, categorical, with the goal of identifying how the information will be processed by the different decision-making systems.

**System boundaries refine definitions of transfer learning.** Transfer learning is a core concept in the science of learning that typically refers to generalizing or adapting prior knowledge to a new context. Despite its importance, transfer is an elusive phenomenon to define and measure (Barnett & Ceci, 2002). These taxonomies often define transfer in terms of surface-level features like the knowledge domain (e.g., math to science, (Barnett & Ceci, 2002)) or context (e.g., theory to applied (Day & Goldstone, 2012; Perkins & Salomon, 1992)). In contrast, how the different perceptual and action-selection systems generalize across conditions is well-studied within neuroscience and psychology (N. J. Cohen & Eichenbaum, 1993; Lakoff, 1990; McClelland et al., 2010; McClelland & Rogers, 2003; Redish, 1999, 2013; Squire, 1987). The instability in definitions and reliance on surface features in studies of transfer learning suggests that a more fundamental unifying taxonomy based on generalization and consolidation could provide researchers a neuroscientific language providing reduced conceptual ambiguity and greater explanatory depth.

First, transfer can occur within each system, but it occurs differently in each system. Perceptual transfer is the generalization of new situations through recognition as modified categories (Hertz et al., 1991; Lakoff, 1990; McClelland & Rogers, 2003), as well as analogical reasoning (Lakoff & Johnson, 2003). Deliberative transfer is the application of familiar choices to new outcomes or new choices to familiar outcomes, occurring primarily through recognition of causal consequences and mental exploration (Bakermans et al., 2025; Buckner & Carroll, 2007; Gupta et al., 2010; Kliegel et al., 2008; Redish, 2016; Samsonovich & Ascoli, 2005). Procedural transfer is the integration of previously automated skills with new automated skills where well-practiced components can be transferred into new sequences (Dezfouli & Balleine, 2012; Graybiel, 2008; Jog et al., 1999; Shadmehr & Holcomb, 1997). Instinctual transfer is an especially fascinating phenomenon which includes transferring group identity (Boehm, 2012; Delplanque et al., 2019; Dziura & Thompson, 2020; McCullough, 2020; Scheepers & Derks, 2016; Wilson, 2002): for example, when a person identifies strongly with their demographic group but not their professional group, then repeated exposure of the instinctual system to evidence and signals that these two groups overlap will lead the person to acquire a sense of belonging with the new group due to transfer of existing group identity.

Second, transfer can also occur between systems. For example, a deliberative plan can enable successful completion of a task that allows the learner to practice the task, thus building up the procedural system skill sets for that task. Likewise, when a procedural routine involves two or more people working together on successful projects, this can lead to a sense of teamwork and the development of community, and learning in the instinctual system. Understanding the mechanisms by which each system learns can allow researchers to more precisely study the important topic of transfer learning.

**Learning outcomes are complementary, not hierarchical**. The multiple action-selection systems perform complementary functions (Redish, 2013). Learning in one system is not categorically better than learning in another system. Theoretical grounding for modern research should account for this complementary nature. Thus, our model suggests proactive caution regarding previous conceptions of human learning and cognition that depict a ranking of higher versus lower mental processes. For example, Bloom's Taxonomy and Dual Process Theories are explicitly hierarchical (Bloom, 1956; Kahneman, 2011; Li et al., 2024). Though a full review of these theories is beyond the scope of this piece, our model stands in contrast to these earlier theories in that our model suggests that each of these decision systems are optimized for different environmental conditions (Mugan & MacIver, 2020; Niv et al., 2006; Redish, 2013). Value judgements drive research design (Campbell et al., 2001; Douglas, 2009), and grounding research on a complementary interaction rather than a hierarchical or competitive picture alters how research questions are framed, how studies are designed, and how results should be interpreted. For example, active research topics including project-based learning, interactive engagement, human-technology interactions, and course structure innovations routinely use the term "higher-order thinking" or "rational decision-making" in the statement of the research question (Li et al., 2024; Yatani et al., 2024). Often, such studies utilize measures and interpret results in a way that discounts the role of the Procedural System in directly supporting the flexible and imaginative function of the Deliberative System. We emphasize that the function of the Procedural System is not just to allow people to perform in special circumstances that require rapid decision making; rather, it allows people to perform complex action sequences in any circumstance that requires integration of multiple skills. By replacing a hierarchical picture with a complementary one, researchers can utilize measures and interpret results that explicitly account for the essential role of the Procedural System in learning high-complexity tasks. With the growing interest in studying students' reasoning, creativity, and critical thinking (Almulla, 2023; Hačatrjana & Namsone, 2024; Park et al., 2023), we argue that it is important never to model the Deliberative System as a "rational" system governing "higher-order" thinking as this explicitly misrepresents the information processing operations of the Deliberative System and implicitly misrepresents the role of all three other systems.

Our model also stands in contrast to the notion that rapid systems contain only "lower" knowledge while slow systems contain "higher" understanding and in contrast to notions of an internal conflict between "rational" versus "emotional" decision-making. While our model suggests that the Instinctual System releases affective states (J. E. LeDoux, 1996), the Deliberative System in our model also utilizes emotional components that are essential to its overall function (A. R. Damasio, 1996; Ledoux, 2000; Redish, 2016; Shenhav, 2024), and neither of these systems is any more rational than the other (A. Damasio, 1994; Redish, 2013). While Dual Process Theories conflate affect with irrationality (Freud, 1923; Gigerenzer & Brighton, 2009; Kahneman, 2011), the multiple decision-system model reviewed here rejects this notion, which is more compatible with leading theories used to study important affective phenomena within formal education which depict a constructive interplay between affect and cognition (Lerner et al., 2015; Pekrun, 2006; Wigfield & Eccles, 2000). This provides a ripe opportunity to more accurately account for the specific role of affect within the Deliberative System (Shenhav, 2024) and the role of probabilistic logic of the Instinctual System (Gallistel, 1990; Rescorla, 1988). One area of active research involving the interplay of affect with cognition is *wise interventions (Walton, 2014)*. Wise interventions are designed with strong theoretical grounding on specific psychological effects. Yet the benefits of these interventions are difficult to replicate and tend to be highly context-specific (Hecht et al., 2023; Parlak et al., 2025; R. A. H. Wang et al., 2017). Likely because the psychological effects themselves tend to be difficult to replicate and highly context-specific (R. A. Klein et al., 2018). A new generation of "wise interventions" grounded in fundamental neural mechanisms would likely be more generalizable and replicable. Since such interventions aim to yield long-lasting effects from a single event, the aforementioned transfer learning phenomena of perceptual reframing and instinctual identity transfer are strong candidates for a new generation of intervention, but with two caveats. First, the theoretical grounding should utilize the probabilistic logic by which the Instinctual System calculates the selection of affective states. Second, since the neural underpinnings of single-event learning depend on extreme events (Perl et al., 2023; Talarico et al., 2004), fitting memories into well-established schemas (Morris, 1981; Tse et al., 2007), and are subject to drift in what is remembered over time (Talarico & Rubin, 2003, 2007) , researchers may need to consider a shift in strategy to multiple repeated interventions when reliable behavior change is a high priority (Domjan, 1998; Gallistel, 1990).

**Universal similarities highlight individual differences.** Just as teaching practices are significantly improved by accounting for individual differences (Hattie, 2008; Hecht et al., 2023; Tomlinson, 2014), accounting for individual differences across learners is one of the most vital aspects of research in the science of learning. Yet there is also extraordinary regularity across all learners (Koedinger et al., 2023), and many individual-level variables show a consistent pattern of having no measurable influence on the effects of an intervention or a learning mechanism (Kizilcec et al., 2020; Kraft, 2020; Simpson et al., 2024; Yeager et al., 2019). This is not surprising since these fundamental information regularities in the brain occur universally across people, context, and action, even though individual differences can alter specific aspects of those processes, access to, and engagement with those mechanisms. This information-based perspective provides new insights into individual-level characteristics with neural underpinnings (i.e. neurodiversity) and provides a new research path to talk about creating external conditions that can accommodate that neurodiversity (Redish & Gordon, 2016; Rust, 2025; Sarrett, 2018; Shaw, 2011).

The universality of the present model does not discount individual differences. Instead, it provides an opportunity for additional rigor in understanding and measuring individual differences. Consider the essential research practice of measuring variation in learners' prior knowledge (Bittermann et al., 2023; Schneider & Simonsmeier, 2025). Typically, prior knowledge is measured as a single index for a given domain of knowledge (music, math, etc.), or parsed out as topics and concepts within that domain (Binder et al., 2019). If, instead, indices for prior knowledge were parsed out based on deliberative versus procedural knowledge, our model suggests that it would allow for more precise measurement of learning effects within those respective systems. Likewise, our model suggests that research on perceptual phenomena such as problem recognition could measure learning more precisely by measuring prior knowledge in the learners' Perceptual Systems. For important questions regarding learners' identity and belonging, some researchers have already begun to use measures and pre-measures that parse out perceptual phenomena from action-selection phenomena (G. L. Cohen & Garcia, 2008; Murphy et al., 2007). In so doing, these studies have yielded models of how learners perceive culturally-relevant threats in classroom environments that are consistent with current models of how people perceive all kinds of threats in any environment. Thus, we predict that their findings are more likely to generalize and replicate. We suggest that using deliberative, instinctual, and procedural categories to parse out the action-selection component of the measures in studies like these would likewise continue to augment their rigor, generalizability, and replicability.

# Limitations and Conclusion

The key contribution of our model lies in the fact that it is a fundamental model derived from first principles and basic science, not applied science. This is also its primary limitation. While this model applies to all people and all subject matter in all contexts, it should be combined with appropriate applied science and practitioner wisdom regarding the individuals, subjects, and contexts involved. This is especially critical when implementing evidence-based learning and teaching strategies, which can fail when neglecting the mechanism by which they work, or when neglecting practical constraints in the context of implementation (S. Allen et al., 2024; Cook et al., 2019; Powell et al., 2015; Soicher et al., 2020; C. Wieman, 2017). Our model empowers educators and researchers to identify and safeguard the mechanisms underlying proven strategies when adapting them to new individuals, subjects, and contexts; and we strongly encourage using our model with due diligence to relevant applied findings and practical expertise.

Learning is a major driver of nearly all human behaviors and is thus an important phenomenon in a vast array of human relationships and institutions. The model of the interaction of the learner with their environment, the system boundaries between action-selection systems, and the information processing operations these systems perform is well-established and has been independently developed by multiple sub-disciplines of neuroscience. Understanding learning as a multi-system phenomenon thus carries implications for domains as varied as education, healthcare, governance, and organizational resilience. By situating our model at the intersection of neuroscience, psychology, and the social sciences, we highlight a framework that can guide both basic research and applied interventions in high-stakes contexts. We hope that our translational work here demonstrates how a modern neuroscientific model of learning can serve as a mature theoretical foundation for research and practice in any context where human learning occurs.